# Cognitive algorithms and systems of episodic memory, semantic memory and their learnings

**Qi Zhang**

qizhangsensor@gmail.com

Sensor System, Madison, WI, USA

**Abstract** Explicit (declarative) memory, the memory that can be "declared" in words or languages, is made up of two dissociated parts: episodic memory and semantic memory. This dissociation has its neuroanatomical basis—episodic memory is mostly associated with the hippocampus and semantic memory with the neocortex. The two memories, on the other hand, are closely related. Lesions in the hippocampus often result in various impairments of explicit memory, e.g., anterograde, retrograde and developmental amnesias, and semantic learning deficit. These impairments provide opportunities for us to understand how the two memories may be acquired, stored and organized. This chapter reviews several cognitive systems that are centered to mimic explicit memory, and other systems that are neuroanatomically based and are implemented to simulate those memory impairments mentioned above. This review includes: the structures of the computational systems, their learning rules, and their simulations of memory acquisition and impairments.

## 1 Introduction

Memory is a single term referring to a multitude of human capacities. Although a universally accepted categorization scheme does not exist, human memory can be divided into short-term memory (working memory) and long-term memory. Long-term memory can be fractionated into explicit (declarative) memory and implicit (nondeclarative) memory (Graf & Schacter, 1985). Implicit memory encompasses priming, perceptual learning, and procedural skills, etc (Squire, 2004), and explicit memory can be further divided into semantic memory and episodic memory (Tulving 1983).

Episodic memory is defined as memory for events; one must retrieve the time and place of occurrence in order to retrieve the event, as in answering the question, "What did you do this morning?" The retrieval query specifies the time, but in order to recall the events, the person must retrieve the place where the events occurred. Semantic memory refers to relatively permanent knowledge of the world, or factual knowledge. Our knowledge that bird has wings, that fire burns, and that Lance Armstrong is a cycling legend, constitutes our factual knowledge or semantic memory.

One idea about the relationship between episodic and semantic memory is that repeatedly experienced events may become represented in a decontextualized form in semantic memory. For example, one should come up with a conceptual knowledge that all birds have wings, after seeing ducks, chickens, robins, crows, and hummingbirds, etc. One should be able to answer the question "Do all birds have wings?" without having to retrieve any specific episode (a particular time and place) in which one encountered a bird.

The relationship between the acquisition of semantic memory and episodic memory is twofold. One is that episodic memory may be a prerequisite for semantic memory. Empirical studies often reveal that amnesic patients, who lost their capacity to retain episodic memory, become almost impossible to acquire new semantic knowledge (e.g., Squire and Zola, 1998), but their previously acquired semantic memory may still be preserved (e.g., Cohen & Squire, 1980). The other aspect is that semantic memory (factual knowledge) is most likely abstracted and generalized from stored past experiences (i.e., episodic memories) through a cognitive process, named memory consolidation.

Besides the difference in properties, episodic memory and semantic memory are also associated with different cortical regions in the brain. It is generally agreed that semantic memory is associated with general neocortex, while episodic memory mainly with the medial temporal lobe (MTL), which includes the hippocampus and its surrounding cortices (e.g., Eichenbaum, 2004). Lesions isolated within the MTL, especially the hippocampus, always lead to various amnesias. For example, a patient who lost his entire hippocampal function would not remember if he had had breakfast or recognize a person he had spoken to minutes ago. However, the same patient was still able to live his daily life and make intelligent conversations with semantic knowledge acquired before his onset of amnesia (e.g., Scoville & Milner, 1957).

Given the importance of the hippocampus to semantic learning, how episodic memory is stored and activated becomes one of the central issues of memory formation. The hippocampus is characterized with sequential learning and spatial navigation capacities (e.g., Levy, 1989; Levy, 1996; Granger et al., 1996; Wallenstein et al., 1998; McNaughton & Morris, 1987), which allow us to retrieve a specific episode with particular sequence of time and coordination of space. Hippocampal cells that are activated when studied subjects (human and other mammals) perform given tasks, are reactivated when the subjects are asleep, especially in rapid eye movement (REM) stage of sleep, which is considered the sign of dreaming. In some studies (e.g., Pavlides & Winson, 1989; Schwartz, 2003; Zhang, 2009a), the hippocampal reactivation during dream sleep is considered the reactivation of episodic memory and is part of the process of memory consolidation. In dream sleep, episodic memory is likely activated in the form of segments instead of whole episodes (e.g., Fosse et al., 2003), and the segments are likely activated randomly, based on the conclusion that dreaming is the result of random impulse (Hobson & McCarley, 1977; Foulkes, 1985; Wolf, 1994).

Furthermore, the information pathways between the hippocampus and neocortex may provide some important clues for how long-term memory is organized and how semantic knowledge is acquired. The hippocampus has a major input pathway from the neocortex (i.e., the entire spectrum of sensory modalities and multi-modal association areas) to the perirhinal and parahippocampal cortices, to the entorhinal cortex, and to the dentate gyrus. And, the hippocampus also has a major output pathway from the subiculum to the entorhinal cortex, and back to the neocortex (e.g., Aggleton & Brown, 1999; Gluck, et al., 2003).

The properties of episodic memory and semantic memory, and their relations, are briefly introduced above. A cognitive system in mimicking human memory and learning, with regards to explicit memory (both episodic memory and semantic memory), is expected to reflect all these aspects. In the following sections, several computational systems are reviewed, and a new approach is then described in detail.

## 2 Computational systems of episodic memory, semantic memory and their learnings

Sun (2004) puts forward four essential criteria for the architecture of cognitive system: ecological realism, bio-evolutionary realism, cognitive realism, and eclecticism of methodologies and techniques. The central point of the realisms is that, in architecturing such a system, we should "aim to capture the essential characteristics of human behavior and cognitive processes, as we understand them from psychology, philosophy, and neuroscience". He further specifies the essential characteristics as bottom–up learning, modularity (specialized and separate cognitive faculties), dichotomy of implicit and explicit memories/processes, synergistic interaction of the two memories/processes. In other words, these characteristics are centered on how memories are acquired, stored, utilized and how memories may interact. The author would like to extend these essential criteria and characteristics to any computational system of either cognitive or connectionist modeling that is aimed to mimic human cognition.

### 2.1 Cognitive systems of learning and memory

Many cognitive architectures have been proposed, including Collins and Quillian's Model (Collins & Quillian, 1969), ACT-R (Anderson, 1983), SOAR (Newell, 1990), EPIC (Meyer & Kieras, 1997), PRODIGY (Minton, 1990), DEM (Drescher, 1991), COGNET (Zachary et al., 1996), and CLARION (Sun et al., 2001), etc. Each of the architectures is essentially a system of learning and memory, regardless what cognitive tasks it may perform. In the following, we will review some of the architectures and compare them against the essential criteria and characteristics given above.

It is noted that, in all these architectures/systems, explicit/declarative memory only refers to semantic (symbolic) memory, and episodic memory has rarely been considered (Sun, 2004). The lack of episodic memory is clearly one of the short-comings existing in these cognitive systems, given the fact that episodic memory is the prerequisite of semantic knowledge as described previously. It is also noted that not all of the cognitive systems are symbolic architectures, and some of them are symbolic-connectionist hybrid systems, e.g., CLARION.

### 2.1.1 Collins and Quillian's Hierarchical Network Model

Collins and Quillian's Hierarchical Network Model has been one of the most influential models in the symbolic approach of memory research. This model is a structure of symbolic knowledge (knowledge tree) that is thought to reflect how people represent and retrieve semantic information, and allow for inferential reasoning, thus, a computational system may be able to comprehend human language.

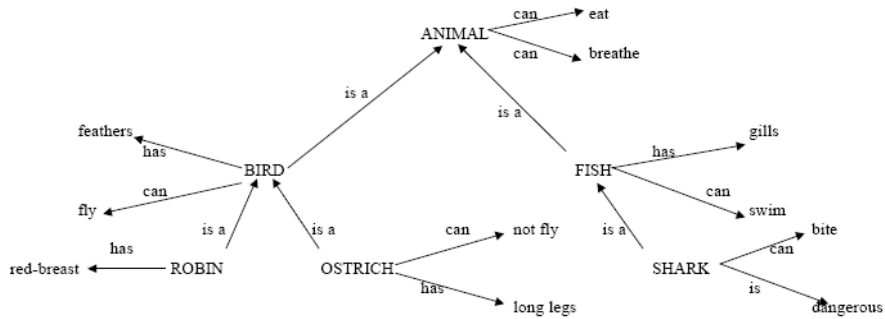

**Fig. 1.** Average verification time for true sentences as a function of number of levels in the hierarchy separating the subject and predicate terms. (After Collins and Quillian, 1969)

The hierarchical network of semantic memory structure is a network of concepts, as shown in Figure 1. The concepts are connected together by labeled relations. The meaning of a concept in this system is therefore represented by the total configuration of relations it has to other concepts. The relations have two kinds: one is category membership labeled by "*is a*", and the other is property relation labeled by "*is*", "*has*" and "*can*". The organization of the network is hierarchical in both membership and property relations. Category members have a direct "*is a*" link to their immediate superordinates (e.g., *robin* is directly linked to *bird*), and properties are only stored at the highest concept level to which they apply (e.g., "*can eat*" is stored with *animal*, rather than with *bird*).

When receiving a statement (e.g., *A robin can fly*), the structure can verify if the statement is true or false, by first entering the network at the node corresponding to the subject term and then searching for the predicate term. The search process first examines the relations that are directly lined to the subject term. If the predicate is found, the search stops and the subject responds "*true*". Otherwise, the search process moves up the hierarchy to next level and examines those relations.

This structure of semantic knowledge is one of the earliest attempts and has outlined the foundation for subsequent developments of symbolic memory structure in many other models. Since this model is not a fully developed cognitive architecture, it is not compared with the four criteria and characteristics. However, there are two issues need to mention. One is that there is no learning mechanism implemented, and as a result the system can only respond and cannot learn. The same issue exists in a few other cognitive architectures, such as PRODIGY (Minton, 1990). The other is that in the semantic network, the relations are hand-coded. The same issue exists in many current semantic networks in which conceptual hierarchies require *a priori* determination through hand-coding, and slots need to be determined also through hand-coding.

## *2.1.2 ACT-R*

ACT-R (Adaptive Control of Thought—Rational, Anderson, 1993; Anderson & Lebiere, 1998) is arguably the most successful cognitive architecture in existence. The model has been applied in capturing a variety of human data in many different task domains, including, e.g., simulations of primacy and recency effects of working memory (Anderson, et al., 1998; Anderson & Matessa, 1997), and modeling of language acquisition and understanding (Anderson, et al., 2004; Budiu & Anderson, 2004), etc. The core of the ACT-R modeling is it's learning and processing algorithm of declarative memory. Based on this central algorithm, several versions of ACT-R system have been developed for various applications with added modules that are intended to match human anatomical cortical areas for executive reasoning, visual perception, and motor control, etc.

Figure 2(a) is the architecture of ACT-R 6.0 of the latest version in which the cognitive center is the combination of the "Declarative module" for declarative memory and "Productions" module for procedural memory. In this modeling, declarative memory consists of symbolic facts such as "Washington, D.C. is the capital of United States" or "3+4=7", and procedural memory is made of production rules, which is intended to mimic human implicit memory about how we do things such as driving car or writing words. The procedural module serves as a switchboard function and connects every module through a designated buffer.

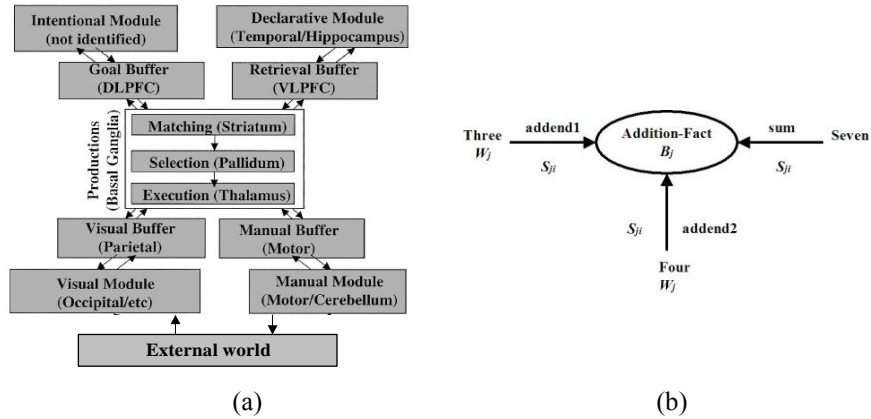

(a)                                                                    (b)

**Fig. 2**. (a) The structure of ACT–R version 5.0. Semantic knowledge (chunks of pointers or symbols) is stored in the declarative module; the associations among the pointers are stored in the procedural (productions) module. (b) A chunk encoding the fact that 3 + 4 = 7. after Anderson, et al. (2004).

According to ACT-R, declarative knowledge is represented in terms of chunks of different types, and each type has an associated set of pointers encoding its contents; procedural knowledge is represented by production rules that are "procedurals" by which the pointers of an associated chunk (declarative knowledge) can be reassembled. Therefore, learning involves both declarative and procedural knowledge (i.e., both chunks and rules). Figure 2(b) is a graphical display of a chunk that encodes "3+4=7" with pointers to *three* ($B_i$), *four* ($W_j$), and *seven* ($S_{ji}$). The production rule of this addition is: IF the goal is the process a string containing digits d1 and d2, and d3 is the sum of d1 and d2, THEN set a subgoal to wrote out d3. Each production consists of a condition (IF) that consists of a specification of the current goal, and an action (THEN). When this learning occurs, a chunk containing the three pointers is stored in the declarative module, and the production rule is stored in the procedural module. The retrieval of a declarative knowledge is driven by production rule. This rule is excited by input from external world and then applied to the declarative module through its buffer to activate and retrieve a specific chunk. The retrieved elements (pointers) of the chunk are "reassembled" in the manual module based on the production rule.

It can be seen that ACT-R modeling does not match the four criteria and characteristics very well. Two important issues are elaborated in here. One is about the problematic division between declarative and procedural knowledge. For example, the association among pointers (e.g., "d3 is the sum of d1 and d2") is not only about how the pointers may be manipulated, but also how we understand the world. Without the association, the pointers are meaningless symbols. In other words, the combination of declarative and procedural knowledge in ACT-R refers to semantic memory of declarative memory of general understanding, and the pro-cedural knowledge in ACT-R is not the procedural memory (like how to drive or how to ride bicycle) of general understanding. This issue leads to the second issue that ACT-R does not address bottom-up learning, because the production rules are only abstracted at top-level. For example, the production rule acquired from the semantic input, "London is the capital of UK", is a manipulation rule at top-level.

### 2.1.3 CLARION

CLARION stands for Connectionist Learning with Adaptive Rule Induction ON-line (Sun, 1999; Sun et al., 2001). CLARION is a hybrid system with a combination of localist and distributed representation. It has a dual-representational structure and consists of two levels: the top level captures explicit processes/knowledge and the bottom level implicit processes/knowledge (see Figure 3a and b). In CLARION version 1, the action and non-action-centered explicit (or implicit) representations are combined in one block (Sun, et al., 2001). Different from existing models of mostly high-level skill learning that use a top-down approach (i.e., turning declarative knowledge into procedural knowledge through practice as reviewed in ACT-R modeling), CLARION uses a bottom-up approach toward low-level skill learning, where procedural knowledge develops first and declarative knowledge develops later.

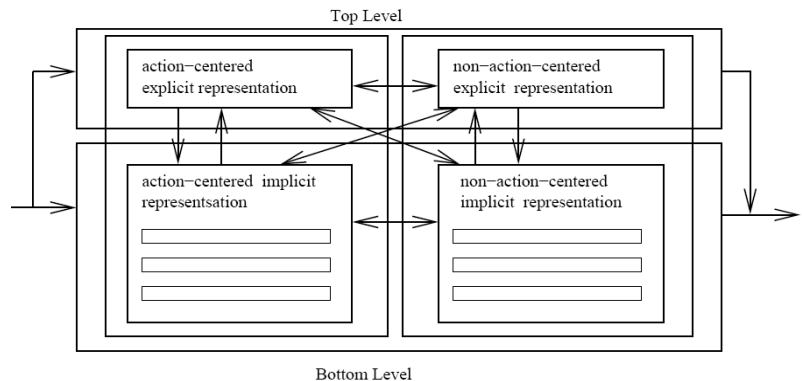

**Fig. 3.** The CLARION architecture. After Sun, 2004.

In the bottom level, the learning algorithm is called Q-learning-backpropagation algorithm, which is a supervised and/or reinforcement learning algorithm adopted from Q-learning algorithm (Watkins, 1989). Such learning is to acquire Q-values. Each Q value is an evaluation of the "quality" of an action in a given state. During learning, Q values are gradually tuned to enable reactive sequential behavior to emerge in the bottom level. The calculation of Q-values for the current input with respect to all the possible actions is done in a connectionist fashion through parallel spreading activation. In the system, a four-layered connectionist network is used in which the first three layers form a backpropagation network for computing Q-values and the fourth layer (with only one node) performs stochastic decision making. The output of the third layer indicates the Q-value of each action (represented by an individual node), and the node in the fourth layer determines probabilistically the action to be performed based on the Boltzmann distribution, given as follows

$$p(a \mid x) = \frac{e^{Q(x,a)/\alpha}}{\sum_i e^{Q(x,a_i)/\alpha}},$$

Here $\alpha$ controls the degree of randomness of the decision-making process (Watkins, 1989), and $x$ is a state of the network.

In the top level, declarative knowledge is in a simple prepositional rule form that captures a bottom-up learning process by using information generated in the bottom level. The correlation between top-level rule and bottom-level output is a set of preset correspondences based on a localist connectionist model with which a set of rules is translated into the network. Assume that an input state $x$ is made up of a number of dimensions (e.g., $x1, x2, \ldots, xn$). Each dimension can have a number of possible values (e.g., $v1, v2, \ldots, vm$). Rules are in the following form: *current-state* 3 *action*, where the left-hand side is a conjunction of individual elements each of which refers to a dimension $xi$ of the (sensory) input state $x$, specifying a value or a value range (i.e., $xi \in [vi, vi]$ or $xi \in [vi1, vi2]$), and the right-hand side is an action recommendation $a$. The top-level learning algorithm is as follows: If an action decided by the bottom level is successful then the agent constructs a rule (with its action corresponding to that selected by the bottom level and with its condition specifying the current sensory state), and adds the rule to the top-level rule network.

The fundamental difference between CLARION and ACT-R is the plausible bottom-up learning algorithm in which a connectionist network is expected to learn implicit knowledge, which becomes the bases of declarative knowledge. This model is a result of the effort to resolve the fundamental and long-standing problem of symbol grounding (Harnad, 1990; Searle, 1980; Smolensky, 1997) by connecting symbols to their meanings that may be acquired by neural network, or by associating rule based knowledge to similarity based knowledge (Sun, 1995). However, in practice, this architecture has not simulated as many human behaviors and cognitive processes as ACT-R has, and the effectiveness of symbol grounding remains to be demonstrated in terms of robustness and flexibility in using acquired knowledge. It is noted that the original version of CLARION has a subsystem of episodic (or instance) memory to store recent experiences in the form of "input, output, result" (i.e., stimulus, response, and consequence), but this subsystem is removed in later versions.

## 2.2 Connectionist systems of episodic memory, semantic memory and their learnings

Some connectionist systems are briefly reviewed for three reasons. One is that, as we have seen, cognitive architectures generally ignore episodic memory, even if episodic memory is the prerequisite of semantic memory/knowledge. The lack of episodic memory is clearly against the essential criteria of cognitive system given by Sun (2004). On the other hand, episodic memory is broadly considered and implemented in connectionist memory systems. The second reason is about symbol grounding. Many believe that, in order for a cognitive system to be robust and flexible, the system has to learn the meanings of symbols. Symbolic-connectionist hybrid system is the best candidate, and neural network is expected to capture the meanings (e.g., Harnad, 1990; Sun, 1995). Thus, one may want to know the progress in capturing meanings through connectionist memory systems. Finally, these systems are able to simulate cognitive process and behaviors of memory consolidation and amnesias, which are also simulated by a cognitive system developed by the author (to be introduced later).

The three connectionist systems (McClelland et al., 1995; Murre, 1996; Meeter & Murre, 2005; O'Reilly et al., 1998; Squire & Alvarez, 1994) to be reviewed are considered neuroanatomically based systems as noted by Meeter & Murre (2005) because of the structural similarity to human brain. They all have the same view that the hippocampus and neocortex play distinct, but complementary, roles in long-term memory (i.e., episodic vs. semantic memory, and fast learning vs. slow learning). They all are able to simulate memory consolidation and retrograded amnesia, and the simulated results are examined by cued recall. Cued recall is one of the standard tests in human memory study. In the test, the subject is firstly presented with an information pair (i.e., picture-word pair), and then is asked to recall the word when promoted by a cue (i.e., the picture).

The system, presented by Alvarez & Squire (1994), consists of two "cortical" areas that are reciprocally interconnected with the MTL area, and the proposed MTL (representing the hippocampus and its surrounding areas) is a temporary connection that binds two separated cortical areas of the proposed neocortex (see Figure 4). Each of the cortical area is made up of two groups of four simplified neurons, whereas the MTL consists of four neurons. During the training phase of episodic learning, the two cortical areas store externally presented patterns, and the MTL stores the "indexes" of the patterns stored in the neocortical areas. The "index" is proposed to point to relevant neocortical cells and activate them (Teyler & DiScenna, 1986) into specific pattern that has been learned in episodic learning. Under the guidance of the indexes, cortical-cortical connections associated with the stored patterns are slowly strengthened during memory consolidation. Simulations show that the system is at chance to perform cued recall when the MTL unit is lesioned soon after the training phase. With sufficient memory consolidation, however, the same damage no longer affects the recall because the connections among the stored information have been established and the binding function of the MTL is no longer necessary. During such tests, the connections between the MTL and cortical areas can be lesioned (inactivated) or normal (kept active). The former is called "lesioned" state that corresponds to amnesic state, and the latter is called "normal" state. These tests yield two curves of activation error vs. consolidation time. The "normal" curve matches the direction and shape of cued recall performed by healthy people, while the "lesioned" curve matches that of cued recall performed by patients with retrograde amnesia, namely Ribot gradient (1881), or temporal gradient. It is explained that, in amnesic state, remote memory is better retained than recent memory because of longer and more sufficient consolidation.

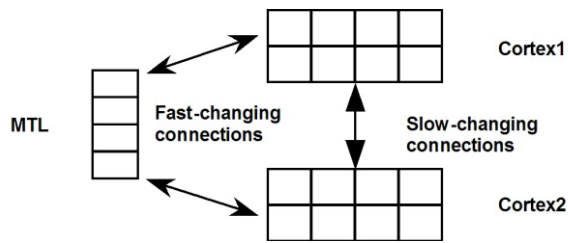

**Fig. 4.** The structure of the binding model (Alvarez & Squire, 1994). Cortex1 and Cortex2 represent two neocortical areas, and each consists of eight neurons. The MTL is made up of four neurons. Each neuron in all three areas is reciprocally connected to each neuron in the other areas. There is no connection within areas, only a form of winner-take-all inhibition.

The TraceLink system, initially proposed by Murre (1996) and further developed by Meeter & Murre (2005), is based on a similar concept as the binding system, i.e., episodic memory is initially stored in the neocortical basis, and consolidation binds the traces of the stored information. The TraceLink system has three subsystems: a trace system (a layer of 200 nodes, i.e., highly simplified neurons) representing neocortical areas, a link system (a layer of 42 nodes) representing MTL, and a modulatory system representing basal forebrain, etc. Similar as the binding model, it is assumed that the formation of associations between neuron groups within the trace system is a slow process compared with the formation between the trace and link system.

This system undergoes four stages for long-term learning. In stage 1, an external pattern activates a set of trace nodes. In stage 2, the activated trace nodes activate a set of link nodes. Stage 3 is considered the initial consolidation process in which the link system is given a burst of random activation to initiate a random search for the nearest representation in the trace system. After a representation is found, the representation remains active until the next burst of random activation in the link system. Consolidation occurs through the formation and strengthening of connections within the trace system at a fixed base rate. In the final stage of consolidation, stage 4, trace–trace connections have become very strong, and retrieval of the initial memory becomes independent of the link system.

Normal learning and temporally graded retrograde amnesia are simulated and tested by meanings of cued recall, and the results are similar as those of the binding model/system reviewed earlier. The simulations of retrograde amnesia is implemented by entirely disabling the link system after initial learning, thus the trace nodes activated by the initial learning in one group cannot form strong associations to the trace nodes in the other group. Anterograde amnesia is the opposite of retrograde amnesia. Patients with pure anterograde amnesia show strong deficit in recalling events experienced after their amnesic onset. TraceLink model simulates such impairment with two kinds of causes. One is a lesioned link system that is similar as in retrograde amnesic simulation. The other is a lesioned modulatory system. As a result, the link system loses its fast learning function and no longer assists the association between groups of trace nodes.

McClelland and colleagues (e.g., McClelland, et al., 1995; O'Reilly et al. 1998) present a different model from the previous two, in which episodic memory is considered to be initially stored in hippocampus. Memory consolidation is considered a "training process", in which the hippocampus slowly teaches the hippocampal representations into the "neocortex". In the simulations of memory consolidation and retrograde amnesia, McClelland et al. (1995) only implement a network system for semantic memory (the proposed neocortex), but not a hippocampal system. The hippocampal functions of rapid learning and information interleaving are assumed through data feeding to the input layer of the semantic system. In the simulations, a three-layer network system (generic three-layered feed-forward network, Mc-Closkey & Cohen, 1989), consisting of 16 input units, 16 hidden units, and 16 output units, is used. The system is first fully trained on a set of 20 random input-output pairs. These pairs are considered as previously acquired experiences. Then, the system continues to be exposed to these pairs throughout subsequent learning of 15 more input-output pairs. Thus, one additional training pair can be "interleaved" into previously learned associations during the new learning. After introduction of the new pair, training continues as before, which is assumed to be consolidation process. And, other new learnings continue in the same fashion for a total of 15 pairs. After all of the new pairs have been learned, the system is examined by being presented with a newly learned input to its input layer at given time intervals of consolidation. The output of the system is compared with the learned output that is assigned to the tested input.

# 3 A multi-leveled network system of episodic memory, semantic memory and their learnings

Human explicit memory (declarative memory) consists of two dissociated components: episodic and semantic memory. Episodic memory is the memory for events that are featured with temporal sequence and spatial coordination of occur-rence, whereas semantic memory is about factual and conceptual knowledge that is independent of a specific past experience. Semantic memory resides in the general neocortex and is independent of the MTL, but newly acquired episodic memory is dependent of the MTL. The development of semantic memory relies on the retention of episodic memory, and semantic memory is most likely acquired from episodic memory in a cognitive process named memory consolidation. Lesioned MTL may lead to various amnesias and result in the acquisition deficit of new semantic memory.

As reviewed earlier, in most cognitive systems, episodic memory is not considered. The lack of the episodic memory indicates the lack of neurobiological realism and cognitive realism in capturing the "essential characteristics of human behavior and cognitive processes". If semantic knowledge has to be consolidated from episodic memory, the consolidation process must selectively consolidate certain information from the episodic memory and ignore others, thus results in the robustness and flexibility of acquired knowledge. Such robustness and flexibility of human knowledge has yet to be demonstrated by computational systems.

On the other hand, the reviewed connectionist systems have fairly captured the relation between episodic memory and semantic memory. However, they fall short in demonstrating the temporal/spatial features in simulated episodic memory, and especially in demonstrating the factual/conceptual properties in the simulated semantic memory. During training phase, the systems are often presented with a series of patterns, but only one pattern, rather than an "episode" of trained patterns (like serial recall), may be recalled at a time. Although the systems are implemented with a semantic subsystem, the recalled materials are those trained patterns, which are arbitrarily patterns and are clearly not factual/conceptual knowledge, despite much hope has been given to neural network in capturing meanings of knowledge (Harnad, 1990; Sun, 1995).

A cognitive system of learning and memory is introduced next (see Figure 6). At structure level, the system is like a typical cognitive architecture in many aspects. It consists of a symbol subsystem and representation subsystem, which are equivalent to the declarative memory and implicit memory in the reviewed cognitive architectures. It employs a bottom-up learning mechanism, which has similar purpose as the sub-symbolic leaning in ACT-R or the similarity learning in CLARION. It will be seen that the system almost agrees with the four criteria and characteristics put forward by Sun (2004). However, there is a fundamental difference. It is not a rule-based system; rather it is a meaning-based system. The bottom-up learning is centered on abstracting and generalizing meanings (common features) from episodic memory. Such a learning mechanism is intended to find a practical solution to resolve the open question of symbol grounding problem. The effectiveness of symbol grounding will be demonstrated.

The presented system also has some similarity to the reviewed connectionist systems. It has cognitive areas that are equivalent to the hippocampus and neocortex, and is able to repeats what have been achieved by the reviewed connectionist systems. In addition, it can also simulate developmental amnesia and direct semantic learning. The same cognitive system and some of these simulations have been reported previously (Zhang, 2005; 2009a; 2009b). The combination of the symbol and representation subsystems makes up the center of the presented system, each of which is a multi-level cognitive construct from base level to subsystem level. This system is introduced in a bottom up fashion i.e., from the base level to the top level of the overall system.

## 3.1 Single memory: to locally store information

A single memory (SM) is the basic cognitive unit of the presented system in this study as shown in Figure 5a, which can store and process either one symbol or one numerical value. A SM has three types of input: signal input ($I_{sig}$), excitation input ($I_{exc}$) and interlock input ($I_{int}$). It has four types of output: signal output ($O_{sig}= I_{sig}$), excitation output ($O_{exc}$), interlock output ($O_{int}$), and coordination output ($O_{cor}$). The signal input is associated with external information and is the subject to learn. Its activation becomes the signal output that is used as a feedback to external world. All other signals are internal signals and are used to organize a dynamic knowledge structure, and to activate the stored external information as well. A SM may learn an incoming signal-input only when it also receives a positive interlock input and an excitation input; it may activate its signal output only when it receives its designated excitation input. The function of the coordination output is to associate a signal-input learned in one subsystem with a signal-input learned in the other subsystem. The learning rules and activation rules are given in Table 1.

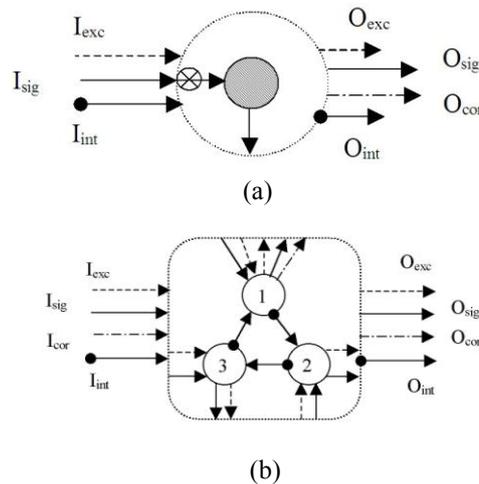

**Fig. 5.** The schematics of (a) single memory and (b) memory triangle, after Zhang, 2005.

A SM is a storage unit to store an *Isig-Iexc* pair, as well as, a comparator that fires its stored information accordingly after comparing an arriving signal with what has been stored. Table 1 indicates that a SM learns in two consecutive steps. In step 1, if condition allows, the SM fires an *Ocor* and waits. In step 2, after the acting SM receives a unique *Iexc* it has been waiting for, it stores the *Iexc* together with the *Isig*. The generation of the unique *Iexc* is a result of the *Ocor*, and this correlation is best explained at system level later.

Memory formation in a biological system is thought to associate with the changes in synaptic efficiency that permits strengthening of associations between neurons, and the synaptic efficiency is related to two phases, short-lived and long lasting, of synaptic modifications. The long-lasting modification may mostly (although not always) be induced by a series of tetanic stimulations over a long period of time in laboratory condition, and is considered an attractive candidate for the molecular analog of long-term memory (see Lynch, 2004). In order to cooperate with the long-lasting modification, a delay, T, is added to the learning mechanism of the SM at Step 2. At step 2, therefore, a SM can fire a positive interlock signal only after it has been stimulated by the same signal input, *Isig* (= *"Io"*), for a given number of times over a period of time.

**Table 1.** Three important states of the single memory in the semantic system, after Zhang, 2009a.

|  | State | Input | Output |
|---|---|---|---|
| Learning | Step 1: firing Ocor signal | Isig = "Io" | Osig = null |
|  |  | Iexc = null | Oexc = null |
|  |  | Iint = "yes" | Oint = "no" |
|  |  |  | Ocor = "yes" |
|  | Step 2: storing "Io" and "Iexco" permanently | Isig = "Io" | Osig = null |
|  |  | Iexc= "Iexco" | Oexc = null |
|  |  | Iint = "yes" | Oint = "yes" |
|  |  |  | Ocor = "yes" |
| Firing stored "Io" upon receiving "Iexco" after the single memory has learned. |  | Isig = any | Osig = "Io" |
|  |  | Iexc= "Iexco" | Oexc = "Iexco" or null* |
|  |  | Iint = "yes" | Oint = "yes" |
|  |  |  | Ocor = "no" |
| Firing stored "Iexco" upon receiving "Io" after the single memory has learned. |  | Isig = "Io" | Osig = "Io" or null** |
|  |  | Iexc = any | Oexc = "Iexco" |
|  |  | Iint = "yes" | Oint = "yes" |
|  |  |  | Ocor = "no" |

\* Depending on Isig
\*\* Depending on Iexc

## 3.2 Memory triangle: to learn meanings or common features

The cognitive capacity of one SM is very limited, and the capacity can be extended when a number of SMs are organized into a group. Three SMs are organized into such a group, named memory triangle (MT), in which three single memories form a loop via interlock signals (see Figure 5b). The function of a MT is to learn a data point ($Isig$) for three times, in case the data point is the only common feature in a number of external representations.

According to Immanuel Kant and John Locke, a concept is a common feature or characteristic, and concepts are abstracts in that they omit the differences of the things in their extension, treating them as if they were identical. In the concept "Bird has wings", "wing" is the common feature of all birds, whereas specific characteristics such as color, size, and sound possessed by a specific bird, can be omitted. The MT is designed to capture a common data point (common feature) and generalize it. In a MT, only one SM is activated to capture the "common data point" at a time. After a MT has stored the "common data point" for three times into each of its SMs in the order from 1 through 3, the common feature is considered learned and generalized because of the existence of the loop. In here the order of learning is regulated by the $Iint$, and the extension of the common feature is realized by the loop.

A loop formed by three SMs, instead of four or five, may be best explained in terms of the principle of minimum potential energy. This principle is one of the fundamental principles we understand about nature. This principle says that a system always intends to configure itself into a formation that has minimum potential energy. The act of the principle is everywhere: the shape of star is always sphere; river runs to ocean; and one oxygen atom bonds to two hydrogen atoms instead of one or three. A loop consisting of three, four or more SMs can perform the same function of common feature extension, but a MT is the smallest loop that requires the least energy to maintain, thus becomes the first choice.

## 3.3 Organizing memory triangles: to learn a knowledge structure

"Knowledge is an integrated phenomenon; every piece of knowledge depends on every other one"; what an intelligent system "has to do is to slowly accumulate information, and each new piece of information has to be lovingly handled in relation to the pieces already in there" (Schank, 1995). Similarly, an external representation may come with only one common feature, but often it comes with more features that may be interlaced and correlated. A number of memory triangles may be organized into a subsystem that can learn more common features that are logically interrelated.

Either one of the two subsystems in Figure 6, the symbol subsystem or representation subsystem, is formed to learn several interrelated common features. In the subsystems, information is locally stored. Where a given external input may be stored is the key for an overall knowledge structure, and is regulated by interlock signals. The rule of interlocking is simple, same as how it works on three SMs within a MT: only when a MT has acquired a common feature, this MT unlocks the next MT. Under this mechanism, *MTr1* must first learn, then *MTr2*, and finally *MTr3*, if the to-be-learned common features are interrelated in a logically hierarchal structure. It learns in a similar way as people do: we have to know the meaning of "zero" before knowing the meaning of "one"; we have to understand what "one" is before knowing what "many" is.

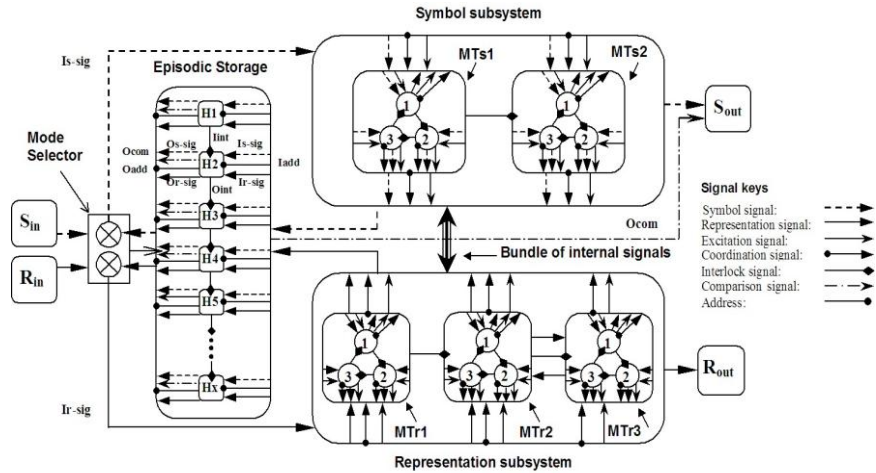

**Fig. 6.** The cognitive structure of a computational system that consists of a semantic system (the combination of the two subsystems) and an Episodic Storage. Each of the subsystem has three layers arranged in a hierarchal order from single memory (each of the small circles marked with 1, 2 and 3) whose function states are given in Table 1, to memory triangle (e.g., *MTs1* and *MTr1*), and to subsystem. The symbol subsystem learns only symbols, and the representation subsystem learns only common features. Conceptual knowledge is acquired when a learned symbol is associated with a learned common feature. The Episodic Storage consists of a number of memory cells whose function states are given in Table 2. All of the cells are interlocked so that the Episodic Storage is able to store and retrieve a sequence of presented events. The Storage can also activate its stored information randomly. During episodic learning, a signal input (*Is-sig*) and a representation input (*Ir-sig*) are presented at *Sin* and *Rin*, respectively. Since the semantic system is a slow learner, it forwards them to the Storage for immediate episodic learning. During sequential recall, the Storage, triggered by the Mode Selector, activates a series of stored events along the interlocked direction. During recognition test, a representation input is presented at *Rin* and is forwarded to the Storage for comparison. A "*yes*" or "*no*" signal of *Ocom* is fired as the result of comparison, and is projected to the *Sout*. During memory consolidation, the stored information is randomly and repeatedly activated from the Storage, and becomes the source of semantic learning of the semantic system. After Zhang, 2009b.

The outline of either the symbol or representation subsystem is called "interface", which is the top cognitive layer of its subsystem and connects all its MTs. An interface is the information gateway of a subsystem, which delivers external stimuli to its MTs, exchanges information between the two subsystems, and projects signal output to the external world and other subsystems. When an interface receives an external signal, it disassembles the signal into a sequence of data points, and distributes the sequenced data points, one by one, to its MTs and SMs. It also collects and organizes activated information and forwards them to the opposite subsystem or external world.

## 3.4 Conceptual learning: to ground symbols to their meanings

A concept is an abstract idea or a common feature, and a word is a symbol for concept. A cognitive system should learn a common feature together with its symbol to complete a knowledge acquisition. It is well known that symbolic approach of cognitive modeling has the advantage in learning symbols, while connection modeling is effective in learning patterns. How to ground symbol to meaning is still an open question. Connectionist modeling can also be developed to learn pure symbols (i.e., the network of symbolic knowledge tree presented by McClelland JL, et al. 1995), however, there has been no substantial progress to ground symbols to their meanings (instead of patterns) in a network. Researchers have made great efforts to answer the question of symbol grounding (e.g., see Sun & Alexandre, 1997). The system presented here can be seen as one of the efforts.

The system illustrated in Figure 6 has two subsystems of symbol and representation, one is dedicated to learn symbols and the other is to learn common features. These two subsystems communicate with each other via the "bundle of internal signals". When the system learns, it abstracts common feature from external representation and stores it in the representation subsystem and does the equivalent to symbol in the opposite subsystem. When learning occurs, the symbol stored in one subsystem is paired up with the common feature stored in the other subsystem. The pairing is realized by excitation input ($I_{exc}$). Excitation input is one of the four inputs of a SM, and is generated by the "bundle of internal signals". The generation only occurs when this bundle receives one $O_{cor}$ from either subsystem. When the acting SM that has fired the $O_{cor}$ in either subsystem receives a newly generated $I_{exc}$, it stores this $I_{exc}$ together with an $I_{sig}$ (see Table 1). Every $I_{exc}$ is unique and is acting like a dynamic "address". So even if every SM in a subsystem is queried by a $I_{exc}$, only the SM containing same "address" can be excited to fire its $O_{sig}$ (see also Table 1).

This combination of subsystems is considered the semantic system of the overall system shown in Figure 6. This combination is inspired by the finding of split-brain (Myers and Sperry, 1953) that indicates each brain half appears "to have its own, largely separate, cognitive domain", and to "have its own learning processes and its own separate chain of memories" as described by Sperry (1982). Sperry further noted that our left hemisphere is capable of comprehending printed and spoken word, and our right hemisphere is word-deaf and word-blind, but capable of comprehending spatial and imagistic information.

## 3.5 Episodic storage: to store episodic memory

The episodic storage, in Figure 6, is a storage site for external information. Due to the modification delay implemented in the SM, the semantic system is unable to learn any external information rapidly, and it always redirects the information to the episodic storage for immediate and direct storage. The direct storage function is same as the notion of "hippocampal system" proposed by McClelland et al. (1995).

**Table 2.** Three function states of the memory cell in the Episodic Storage, after Zhang, 2009b

| State | Input | Output |
| --- | --- | --- |
| Learning: To store "Iso" and "Iro" | Is-sig = "Iso" <br> Ir-sig = "Iro" <br> Iint = "yes" <br> Iadd= null | Os-sig = null <br> Or-sig = null <br> Oint = "yes" <br> Oreco = null <br> Oadd="address-o" |
| Firing: To fire stored "Iso" and "Iro" | Is-sig = null <br> Ir-sig = null <br> Iint = null <br> Iadd = "address-o" | Os-sig = "Iso" <br> Or-sig = "Iro" <br> Other signals = null |
| Comparing: To compare incoming signal with stored "Iso" and "Iro" | When Is-sig = "Iso"; other signals = null | Oreco = "yes"; Or-sig = "Iro"; other signals = null |
| | When Ir-sig = "Iro"; other signals = null | Oreco = "yes"; Os-sig = "Iso"; other signals = null |
| | When Is-sig ≠ "Iso"; or Ir-sig ≠ "Iro"; other signals = null | Oreco = "no"; other signals = null |

The Episodic Storage consists of a number of memory cells (MC) that are enclosed by an interface that delivers inputs in parallel to all MCs and collects outputs from them. Each MC has four inputs (symbol input, *Is-sig*, representation input, *Ir-sig*, interlock input, *Iint*, and address input, *Iadd*) and five outputs (symbol output, *Os-sig*, representation output, *Or-sig*, interlock output, *Oint*, comparison output, *Ocom,* and address output *Oadd*). The function states of a MC are given in Table 2. When a MC learns, it stores a pair of *Is-sig* and *Ir-sig*, and sends its "address", *Oadd,* to the Storage interface. Since all MCs are interlocked by interlock signals in one direction, a sequence of external events can be both stored and activated in the original order of arrival. When a MC receives an external signal that matches any one of the two stored signals, it fires an *Ocom* of "yes", otherwise, "no". The interface can activate the MCs to fire stored signals along the interlocked sequence by sending all MCs a sequence of specific addresses, or activate them to fire randomly regardless of existing sequence. This function is to mimic the sequential learning function of the hippocampus that has been concluded in many studies. The episodic storage receives both symbol and representation inputs from the semantic system as shown in Figure 6, which coincides with the fact that the hippocampus mainly receives inputs from the neocortex [2, 46].

This storage has its designated information pathways to and from the semantic system of the paired subsystems, which coincide with the major pathways concluded in the studies by Aggleton & Brown (1999) and Gluck, et al. (2003).

Other components of the system in Figure 6 are explained as follows. *Sin/Rin* are external input interfaces and *Sout/Rout* are external output interfaces of the system. The Mode Selector is a switch to select input source for the semantic system. The input source can be external information or internal information coming from the Storage during "dream sleep". The Selector can also send a simple triggering signal to the Storage's interface to stimulate sequential or random firing from there.

## 4. Simulating episodic memory, semantic memory and their learnings

The presented multi-leveled memory system is employed to simulate serial recall, memory consolidation, dreaming, retrograde amnesia, developmental amnesia, and direct semantic learning. In the simulations, the episodic memory is demonstrated to be episodic-like, e.g., it may recall an "episode" of past experiences. The semantic memory is demonstrated to be conceptual, e.g., the acquired semantic knowledge can be utilized to process unfamiliar external information.

In the simulations, results are examined in terms of serial recall, recognition, object naming (may also be seen as cued recall), and object drawing. These are all standard tests in psychological studies for learning and memory. It is noted that object naming is similar to, but beyond cued recall. When the system is presented with an experienced input, the result is equivalent to cued recall, but when it is presented with an unfamiliar input, the result is an object naming. Cued recall is almost the only testing method used in the three reviewed computational systems.

## 4.1 Episodic learning, serial recall and recognition

Episodic memory is the explicit memory for events. One must retrieve the time and place of occurrence in order to retrieve the event. The sequential learning and spatial navigation capacities of the hippocampus (e.g., Levy, 1989; Levy, 1996; Granger et al., 1996; Wallenstein et al., 1998; McNaughton & Morris, 1987) play an important part in episodic memory, and allow one to retrieve a specific episode with particular sequence in time and coordination in space.

**Table 3(a).** Input pairs for episodic learning

| Sequence | Symbol | Representation |
|---|---|---|
| 1st | III | 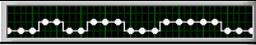 |
| 2nd | I | 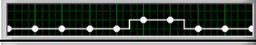 |
| 3rd | I | 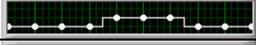 |
| 4th | Z | 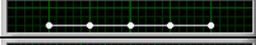 |
| 5th | IIII | 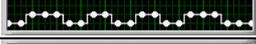 |
| 6th | Z | 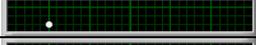 |
| 7th | Z | 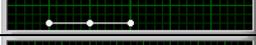 |
| 8th | I | 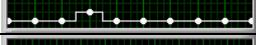 |
| 9th | II | 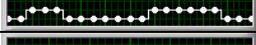 |
| 10th | Z | 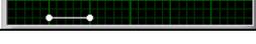 |

**Table 3(b).** Input pairs for direct semantic learning

| Sequence | Symbol | Representation |
|---|---|---|
| 1st | Z | 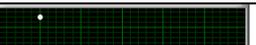 |
| 2nd | Z | 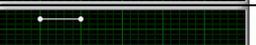 |
| 3rd | Z | 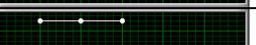 |
| 4th | I | 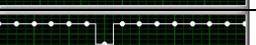 |
| 5th | I | 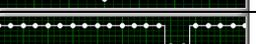 |
| 6th | I | 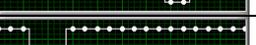 |
| 7th | II | 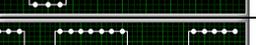 |
| 8th | III | 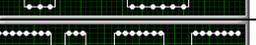 |

The first step in all following simulations is episodic learning, i.e., let the system learn a sequence of external "events". In episodic learning, the Mode selector is set for the system to receive externally presented "events" given in either Table

3(a) or (b). Each "event" in the tables is a symbol-representation pair, and the representation input contains the meaning assigned to the symbol. All input pairs in either Table 1(a) or (b) contain three concepts, "zero", "one" and "tally", which are represented by the common features carried by the representation inputs.

It is noted that the common features for "one" and "tally" are peak/peaks in Table 3(a), and are pit/pits in Table 3(b). In episodic learning phase, we only let the system learn from either Table 3(a) or (b). After the system has gone through semantic learning phase, the system is expected to be able to tally either peaks or pits. The purpose is to show the system's flexibility in learning different common features. For simplicity, however, in most of the simulations to follow, the episodic learning is the sequenced pairs in Table 3(a), and only once the pairs in Table 3(b).

In the episodic learning, each input pair is presented to the system based on the sequence indicated in the table. The symbol input is at the *Sin* and the representation is at the *Rin*. This learning is a one-time experience, and the system is expected to remember the sequenced event thereafter.

The input pairs are firstly transported to the semantic system for learning. However, in most of the cases, the semantic system is not able to learn external information due to the "modification delay" and the complexity of external information. As a result, the input pairs are sent to the Episodic Storage, one after another, for immediate storage into different MCs along the interlocked direction.

The system may be "asked" to recall the sequenced events it has just experienced. During the sequential or serial recall, the Mode Selector sends a trigger signal to the Episodic Storage. In turn, the Storage's interface sends a sequence of *Iadd* to all MCs to activate appropriate memory input pairs. Since the *Iadd* is associated with the interlocked chain of MCs, a past experience is recalled in the same sequence as what has been experienced in episodic learning. The first simulation in Table 4 is such a sequential recall.

**Table 4.** Simulations of sequential recall and recognition after episodic learning

| Task | Input | Output |
|---|---|---|
| Sequential recall | Triggered by the Mode Selector | III I I Z IIII Z I II Z |
| Recognition | 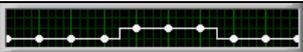 | yes |
|  | 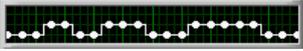 | yes |
|  | 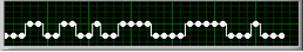 | no |

The system may also recognize the input if a newly presented item is an experienced one. In this process, externally presented information is forwarded to every MC in the Episodic Storage for comparison. When a match is found, a "yes" output is fired from the specific MC. The last three simulations in Table 4 are recognition tests. In these three tests, the first two representation inputs are includ-ed in Table 3(a), and have been "memorized" in the Storage. The system "recognizes" them and shows "yes" to confirm. The last presented input is not included in table 3(a), thus the Storage has no memory of it and fires a "no" for the unrecognized input.

## 4.2 Dreaming, learning and memory consolidation

Dreaming refers to the subjective conscious experience we have during sleep. Numerous studies have concluded that dream deprivation always causes poor mastering of knowledge (explicit memory) and skill (implicit memory) that have been learned in the previous day. Findings of the correlation between dream sleep and waking learning have suggested that dream sleep may play an important role in learning and memory consolidation (e.g., Bloch, et al., 1979; Fishbein, 1970; Greenberg & Pealman, 1974; Pearlman, 1971).

The relation between dream sleep and memory consolidation is also proposed in the studies of neuronal recording, which reveal the replaying of recent waking patterns of neuronal activity within the hippocampus during sleep, especially dream sleep (e.g., Pavlides & Winson, 1989; Wilson & McNaughton, 1994; Staba, et al., 2002; Poe, et al., 2000; Louie & Wilson, 2001). Importantly, this replay, or hippocampal firing, is synchronized with activities in the neocortex, rather than an isolated activity. Such synchronization is attributed to be the evidence of memory consolidation from the hippocampus into the neocortex (Battaglia, et al., 2004).

It is often concluded that dreams are more or less random thoughts, and are caused by random signals (Hobson & McCarley, 1977; Foulkes, 1985; Wolf, 1994). In reviewing the correlation between daily experiences and dream contents, it is found that daily experiences are often replayed in the form of segments, rather than entire episodes during REM (rapid eye movement) sleep (Fosse, et al. 2003).

In short, dreaming may be a learning and memory consolidation process in which the segments of daily experience are randomly activated from the hippocampus, and the neocortex is synchronized to incorporate with the randomly arriving information. Such a process is simulated with the system as shown in Figure 6.

In the simulation of dreaming, the Mode Selector is set for the Episodic Storage to randomly activate its MCs to fire stored information pairs. A stream of activated events flows to the semantic system for further process and learning, and this stream can be recorded as "dream report" (Zhang, 2009a). At the subsystem level, every randomly arriving representation "event", or symbol "event", is disassembled into a sequence of smallest information pieces, and these pieces are, one after another, delivered to every SMs. Each SM may react to, ignore, or learn from the arriving external signal according to the rules given in Table 1. The four levels of hierarchal structure from SM to semantic system regulate whether a SM has a potential to learn. The regulations decide whether an external input fits an existing memory structure, and where the external input should be stored. All of the regulations are simply reflected at the interlock signal inputs among SMs and among MTs. When learning occurs, a symbol is stored in the symbol subsystem and its associated common feature(s) stored in the representation subsystem spontaneously. In the meantime, the stored symbol is paired up with the stored common features by a unique excitation signal that is automatically assigned by the semantic system.

The system has to experience thousands of random activations before it is able to fully consolidate those memorized episodic events into the semantic system. After full consolidation, the system can be set to "waking mode" to process other external information. The last three simulations in Table 5 summarize how the system may respond to external information after the consolidation with a disabled Episodic Storage. It is noted that the three external representation inputs in these three tests are not the ones that have been presented during episodic learning. However, the system is able to count how many peaks exist in the given external inputs. It is able to do so because the semantic system has acquired the common features or conceptual knowledge, and is able to flexibly use the knowledge to process either familiar or unfamiliar information.

## 4.3 Retrograde amnesia and anterograde amnesia

Patients with severe bilateral lesions in the hippocampus, are often unable to remember events from moment to moment, and show a mild loss of old memories extending back in time for years (e.g., Anon., 1996; Scoville & Milner, 1957; Squire & Zola, 1998). The former is named anterograde amnesia and the latter is named retrograde amnesia. Most patients with retrograde amnesia show a temporal gradient (Ribot gradient) in memory retrieval, i.e., episodic memory acquired long before the lesion is better recalled than that of newer memory, which is also named temporal graded retrograde amnesia.

In the three computational systems reviewed earlier, both anterograde and retrograde amnesias are explained in terms of loss of the hippocampal function to bind (Alvarez and Squire, 1994), to link (Murre, 1996) episodic memory that are stored in the neocortex, or to rapidly store episodic memory (McClelland *et al.* 1995). Memory consolidation is exclusively proposed to be the key reason to cause the "temporal gradient" that is observed in most patients with retrograde amnesia. Old episodic memory has more chances than newer memory to be consolidated into the neocortex and to become independent of the hippocampus, thus to be better retrieved in the absence of functional hippocampus. Retrograde amnesia is almost accompanied by anterograde amnesia in cases of the bilateral lesions. So the cause of retrograde amnesia also applies to anterograde amnesia, because without a functional hippocampus, episodic memory after the onset of the bilateral lesions cannot be established and retrieved.

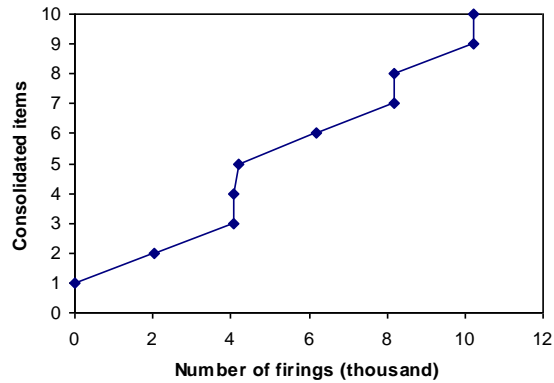

**Fig. 7.** A simulation of memory consolidation showing the relationship between consolidated items and the number of random firings. Here, the ratio of T/t (modification delay/random firing interval) is set at 2000. After Zhang, 2009b.

The key simulation of retrograde amnesia is to exhibit the temporal gradient. Those three computational systems have demonstrated this property under the same mechanism that the temporal gradient reflects the progress of memory consolidation. A similar property is also simulated using the system presented in this chapter, as shown in Figure 7. This temporal gradient curve is a relationship between the number of random activations and number of episodic events that have been consolidated into the semantic system. Since the random firings are activated at a fixed time interval, this curve also represents a relationship between consolidation rate and time. The curve is obtained from scores of cued recall tests, the same method used in the simulations in other studies. In the tests, the cues are those representation inputs listed in Table 3, and the targeted recall materials are those symbols that are paired with the representations. During such a test, the system is set to go through a given number of random activations, then the Episodic Storage is disabled, finally it is presented with a cue and its symbol output at *Sout* is examined.

However, if the Episodic Storage is disabled before episodic learning, the system is unable to recall any of the experienced events, which is a similar mechanism of anterograde amnesia simulated by Meeter and Murre (2005).

## 4.4 Developmental amnesia

Developmental amnesia is an atypical form of memory deficit that has been discovered to occur in children with hippocampal atrophy. A clear dissociation has been revealed between relatively preserved semantic memory and badly impaired episodic memory. Such patients always suffer bilateral damage to the hippocampal formation at very early ages with sparing of surrounding cortical areas. The badly impaired episodic memory is mostly shown in delayed sequential recall and spatial recall. The patients may score anywhere from a few percent up to 25 percent, compared with control groups, in both delayed storytelling and delayed reproduction of geometric designs. However, they have compatible IQs as those of control groups and their recognition ability appears to be normal or close to normal (Vargha-Khadem, et al, 1997).

It seems that such early loss of episodic memory may impede cognitive development and result in severe mental retardation (Baddeley, et al., 2001), since many believe that semantic memory is mainly acquired from episodic memory through memory consolidation. Several explanations have been suggested. One is that the recollective process of episodic memory is not necessary either for recognition or for acquisition of semantic knowledge (Baddeley, et al., 2001; Vargha-Khadem, et al 1997). However, this explanation does not really offer a mechanism for why such patients may still presumably learn semantic knowledge from memory consolidation and recognize presented items, but perform poorly when recalling a sequence of events or spatial related information. Another explanation (Squire and Zola, 1998) is that since none of the patients have entirely lost their "recall memory", the residual "recall memory" may be enough to explain the near normal semantic memory performance, although no detailed mechanism is offered either.

**Table 5.** Simulations of impaired sequential recall, and intact recognition and semantic learning

| Condition | Input | Output | Comment |
|---|---|---|---|
| Tested after episodic learning, but before consolidation | Triggered by Mode Selector | I   Z | Impaired experience recall |
| | 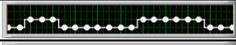 | yes | Intact recognition |
| | 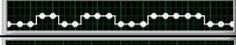 | yes | Intact recognition |
| | 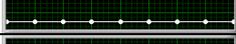 | no | Intact recognition |
| Tested after consolidation and with disabled Storage. | 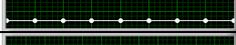 | Z | Intact tally |
| | 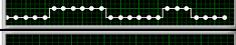 | II | Intact tally |
| | 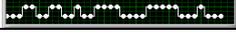 | IIIIII | Intact tally |

While the mechanism of developmental amnesia still remains unclear, the system shown in Figure 6 is the only one that is able to simulate an impaired sequential recall versus an intact capacities of semantic learning and recognition (Zhang, 2009b). These simulations are based on the proposal that limited hippocampal atrophy (27-56 percent compared with healthy subjects, see, Isaacs, et al., 2003) may only impair its sequential and spatial learning capacities, but spare its storage function. It is considered that, in order to memorize sequential or spatial information, a system needs to memorize not only the elements in the information, but also the associations among the elements. When a lesioned hippocampus is no longer able to store the associations, it is problematic to recall the sequential or spatial information entirely, even if every element of the information has been stored. The system may still be able to recognize a past event and learn semantic knowledge from those disconnected events, but is unable to recall them in their originally presented order.

Two mechanisms are implemented to cause the impaired sequential learning or impaired associations of information in the simulations. One is that the interface of the Episodic Storage is unable to register or encode most of the associations, and the other is that most of the associations have been encoded wrongly, for example, incorrect addresses are provided during episodic learning. These two mechanisms imply that the lost associations are fixed at the moment when episodic learning occurs, and are not randomly selected during recall. In other words, the patients may show the same pattern of memorized elements versus lost elements in retelling of the same story and redrawing of the same picture in repeated tests. However in all reported empirical studies, the repeatability is not reported.

The system implemented with either lesioned situation mentioned above is used to simulate developmental amnesia and some of the results are shown in Table 5. The first simulation is a sequential recall, which is apparently an incomplete recall compared with the same recall in Table 4. This partial recall is comparable to the data given in the initial study (Vargha-Khadem, et al. 1997) in which recalled materials are in the range of 20% - 25% of controls in either delayed storytelling or redrawing of presented geometric design. The pattern of the simulated performance is similar to the redrawing of a geometric design performed by three patients in the initial study. In the study, the geometric design is a single structure consisting of many interlaced triangles, rectangles, and lines. The patients are only able to redraw a small portion of the whole design. Interestingly, the redrawn portions are mostly detached triangles and rectangles and the associations among the patterns are lost. The same feature of detached elements is also shown in the simulations of impaired sequential recalls.

The damaged sequential learning mechanism does not necessarily impair the recognition ability of the Storage, because recognition process utilizes the comparison function that is a different mechanism from sequential recall. Information can be recognized as long as it has been stored even with a wrongly registered address. The three simulations show that experienced events can always be recognized (the two recognition tests that generated "yes" output in Table 5), while non-experienced events cannot (the one recognition test that generated "no" output).

Random activation of the proposed hippocampus has also been implemented for the memory consolidation simulations in the three reviewed systems. When semantic knowledge can be learned from randomly activated episodic memory in memory consolidation process, the related semantic learning should be less affected by an impaired sequential learning function. The last three simulations in Table 5 show that the semantic knowledge acquired from randomly activated infor-mation is not only equivalent to, but also beyond what has been learned in episodic learning. In these simulations, the semantic knowledge has been utilized to process "unfamiliar" external inputs even if they are seemingly more complex. This demonstration of semantic knowledge property is considered the basic requirement for the simulation of developmental amnesia, since normal IQ is the key characteristic of the patients with years of developmental amnesia.

## 4.5 Dense amnesia and direct semantic learning

Densely amnesic patients not only show a total loss of episodic learning capacity, but also become almost impossible to acquire new semantic knowledge (Squire and Zola, 1998). When such patients are tested for semantic learning, e.g., new words, over a relatively short period of time (e.g., days or weeks) and with infrequent encounters of learning materials, the results are always negative (Gabrieli et al., 1988; Postle & Corkin, 1998), although the learning conditions are adequate for healthy subjects. The understanding is that such patients are not able to hold new episodic memory, which can be re-accessed for numerous times over a period of time in memory consolidation process to acquire semantic knowledge.

However, the same patients may very slowly acquire semantic knowledge, if they have repeatedly encountered the same information over years of time. The most significant case of the slow semantic learning reported (O'Kane et al., 2004) is about the famous patient, H.M. He is the most studied amnesic patient, and his case has a special position in the understanding of human memory system because of his well-known and well-localized MTL lesion that has left him with no hippocampal function (Scoville & Milner, 1957). In tests, H.M. is able to tell the last names of more than one-third of people who became famous after his amnesic onset, when whose first names are provided as cues. He is able to describe John Glenn as "the first rocketeer" and Lee Harvey Oswald as a man who "assassinated the president". This new knowledge is demonstrated to be flexible and semantic (O'Kane et al., 2004) because H.M. is able to retrieve the same knowledge promoted by different cues. Slow semantic learning, over a long period of time (e.g., 13 years), has also been observed in other densely amnesic patients (Butters, et al., 1993; Kitchener, et al., 1998; Tulving, et al., 1991).

Given the fact of H.M.'s well-known hippocampal lesion, the semantic knowledge he is able to demonstrate is unlikely acquired through the mechanisms identical to the ones that healthy adults use to acquire semantic knowledge. It is suggested that H.M.'s mechanism for semantic learning appears to be via slow learning, whereby following extended and repeated encounters of the same information (O'Kane et al., 2004). Other similar studies (Butters, et al., 1993; Kitchener, et al., 1998; Tulving, et al., 1991) have also come to the same conclusion that the demonstrated semantic knowledge may have been acquired directly and gradually by the neocortex in years of extensive exposures to information. On the other hand, no computational system has been reported previously to simulate the proposed direct and gradual learning mechanism, and more importantly, to demonstrate the learned material is semantic knowledge.

**Table 6.** Naming and drawing after given numbers of repetitions

| R* | External stimulus | Output |
|---|---|---|
| 300 | Z | [drawing] |
|  | [drawing] | Z |
|  | [drawing] |  |
|  | I |  |
| 500 | I | [drawing] |
|  | Z | [drawing] |
|  | [drawing] | I |
|  | II |  |
| 550 | II | [drawing] |
|  | [drawing] | II |
|  | [drawing] | I |
|  | III |  |

R*: number of repetitions

The semantic system shown in Figure 6 can be employed to simulate the direct and gradual process of semantic learning. In the simulations, the Episodic Storage is removed to incorporate with an entirely nonfunctional hippocampus. The learning materials are the external input pairs listed in Table 3(b). The learning procedure is to present the input pairs, one after another, along the given sequence, to the semantic system for a great number of repetitions. After a given number of repetitions, the learning progress is examined in terms of object naming and object drawing.

The first four simulations in Table 6 are the test results after 300 direct learning repetitions, which show that the semantic system is able to use its knowledge about "zero" to process external information, but is unable to understand "one" or "many". When the repetition further progresses, it is able to understand "one", but not "two" after 500 repetitions (the second group of four simulations), and then "two", but not "three" after 550 repetitions (the third groups of four simulations). The representation inputs in the tests are similar in concept to, but different in details from the ones that have been repeatedly presented to the system. The system is able to perceive the meanings from them by flexibly using its acquired knowledge and giving correct answers. One may have noticed that the meanings for tallying given in Table 3(b) are pit/pits, instead of the ones in Table 3(a) of peak/peaks. This new kind of meaning is used for the purpose of demonstrating the relative flexibility of the semantic system in learning different concepts.

### 4.6 Robustness and flexibility

Human knowledge is meaning based, and is robust and flexible. Similar robustness and flexibility has been sought in various computational systems, for example, CLARION can be considered as one of the approaches. It is believed that only when a system is able to acquire meanings from external information, it may exhibit strong robustness and flexibility (Harnad, 1990; Sun, 1995). This presented system is architectured to acquire (abstract and genreralize) meanings from external information. As a result, it has exhibited strong flexibility in using its acquired knowledge in many aspects.

The flexibility is summarized. Firstly, the system can perform variety of cognitive tasks that are often employed in human memory study, e.g., serial recall, cued recall, object naming, object drawing and recognition. Secondly, it can tally any given number of objects and "draw" any number of objects, although it has only learned a maximum of "three" (III) as shown in Table 3(a) or (b). This flexibility demonstrates that the system has truly acquired and generalized the related meanings from given examples. Thirdly, it can tally unfamiliar object that is different from any learned example. Finally, it has certain fuzzy capacity to deal with irregular input. These flexibilities match the criteria outlined by Sun (1995), including *generalization* from examples, *similarity-based* cognition, handling *inexact matches*, and handling *fuzzy* information.

## 5. Future challenges

This multi-leveled network system succeeds in mimicking many properties of episodic memory and semantic memory, and their relationships. It interprets and simulates more phenomena about human episodic and semantic memories and their learnings, than many other reported systems. It suggests a mechanism for the cause of developmental amnesia, and predicts a pattern of forgetting versus remembering in repeated recall tests. In a recent communication with one of the principle researchers who reported development amnesia, it is said that the prediction is most likely true based on some existing data although the patients have not been tested purposely for repeated recalls. On the other hand, this system has been tested with a number of testing methods, such as, serial recall, cued recall, recognition, object naming and object drawing, which are commonly used in the study of human learning and memory. As a comparison, the three reviewed systems may only be tested by one method, e.g., cued recall. When testing method changes, those systems would cease to function as reviewed previously. Furthermore, by grounding symbols to their meanings, this multi-leveled system is able to flexibly use its conceptual knowledge to process unfamiliar information, compared with the reviewed computational systems that only process and recall arbitrary patterns.

Although this multi-leveled system/algorithm has shown a number of promising cognitive capacities, its basic cognitive unit, SM, is not neuron-like. Thoughts have been given for how to make the SM compatible to biological neurons. It is likely that one SM can be formed with a number of artificial neurons of different kinds. Such a possibility is obvious because SM is a generic cognitive unit in the cognitive system, regardless what external information it may associate with, a symbol or a meaning, also because biological neurons are believed to be generic cognitive units in the brain.

The multi-leveled system is able to abstract and generalize a few numerical concepts from given examples, and to tally either peaks or pits from externally presented representations thereafter. A similar system has been trained to learn Arabic numerals that are used as alternatives of the symbols for tallying (Zhang, 2005). How to extend the system to learn non-numerical conceptual knowledge will be one of future efforts, which may involve a number of cognitive aspects. This multi-leveled system is designed to process sequenced information, like a person who may only make sense of the surrounding world by continuously touching. Mechanisms are needed for the system to abstract and learn common features from parallel information (e.g., vision). Fortunately, substantial progresses have been made in visual perceptions that may help to overcome this issue. Furthermore, the spatial learning capacity of the hippocampus may also shed light on this effort. Another aspect is about the boundaries of concepts. Many concepts are true only within given boundaries. We may learn a concept from positive examples, and we may also learn its limitation from negative examples. This multi-leveled system is only able to learn conceptual knowledge from positive examples. Thus, further development is also needed in the respect of concept boundaries.